\documentclass[12pt]{article}
\usepackage{amsmath}
\usepackage{amsfonts}
\usepackage{amssymb}
\usepackage{dsfont}
\usepackage{mathrsfs}
\usepackage{graphicx}

\usepackage[body={16.5cm, 22cm},right=2.2cm]{geometry}

\newcommand{\luv}{\Lambda_{UV}}
\newcommand{\ie}{\textit{i.e.}~}
\newcommand{\xx}{\mathrm{x}}

\begin{document}
\begin{center}
\begin{flushright}
AEI-2009-103
\end{flushright}
~
\vspace{1.cm}

{\Large
{A note on particle kinematics in Ho\v{r}ava--Lifshitz scenarios
}}\\[1cm]

Lorenzo Sindoni
\\[0.5cm]

{\small \textit{Max-Planck-Institut f\"ur Gravitationsphysik \\
Albert-Einstein-Institut\\
Am M\"{u}hlenberg 1,
D-14476 Golm,
Germany}\\
{e-mail: \texttt{sindoni@aei.mpg.de}}}

\end{center}

\vspace{.8cm}
\vspace{0.3cm} {\small \noindent \textbf{Abstract} \\
\noindent In this short note we clarify a link between anisotropic-scaling scenarios and Finsler spacetimes.
Generalizing earlier analysis it is shown that the kinematics of propagating particles (in the sense of
geometrical optics) can be described in terms of (pseudo-)Finsler structures.}

\vspace{1cm}

In recent years, the possibility of a UV-completion of the standard model encoding some form of Lorentz
 violation \cite{LIVSM} has been discussed in different frameworks and from different points of view. In fact, 
it turns out that rather diverse quantum/emergent gravity
scenarios are providing some hints that such an option deserves some exploration \cite{LIV,Emergent} (see also \cite{VisserLIV}). 
In parallel, an increasing number of data coming from (astro-)particle physics experiments have made possible
to provide constraints for some of these frameworks encoding Lorentz symmetry violation.
For reviews and complete 
lists of references see for instance \cite{GAC,LiberatiMaccione}.   

More recently, it has been proposed that scenarios based on anisotropic scaling \cite{Horava} could be
 helpful in providing a coherent setting for a renormalizable quantum theory of gravity, in an appropriate
 sense. In this setting, four dimensional spacetime is described in terms of a preferred foliation,
 $\mathcal{M}=\mathbb{R}\times \Sigma$, with respect to which the notion of anisotropic scaling is
 introduced. For the general presentation and more details see \cite{Horava,HoravaMembranes}.

One of the most important aspects to be understood is the way in which the kinematics of particle physics 
is changed in such a scenario. This problem is just a generalization of the issue of (Lorentz-violating)
 Modified Dispersion Relations (MDR) considered in the past for the standard model alone, with gravity
 being switched off \cite{MyersPospelov} (see also \cite{Pavlopoulos} for an earlier proposal).  

In Ho\v{r}ava--Lifshitz scenarios, the geometry of spacetime where the various fields are propagating is no more a
flat, non-dynamical structure, but rather it satisfies some equations of motion, which should reduce to the familiar
Einstein equations for a four dimensional metric $g_{\mu\nu}$ on large distances. Besides the ongoing analysis
of the viability of such a scenario \cite{HoravaDevel}, 
it is interesting to see what are
the consequences for matter fields, and, in particular, the way in which the propagation of signals is altered with
respect to local Lorentz invariant models \cite{FieldLifshitz}. 

The aim of this note is to put in relation this new direction of research with past work on Lorentz-violating extensions of the standard model, in particular in giving to the MDR appearing in
this new framework a four dimensional geometrical interpretation. We will follow closely \cite{GLS},
and we refer to this paper for a careful discussion of all the technical details.

As an illustrative model, we will use the case of a single scalar field. We will comment later on the case 
of higher spins.
Following \cite{FieldLifshitz}, we can write the most general action for a scalar field as
\begin{equation}
 S_{0}= \frac{1}{2} \int d^4x \sqrt{h} \mathcal{N} \left\{ \frac{1}{\mathcal{N}^2} \left( \partial_t \phi 
-\mathcal{N}^i \partial_i \phi \right)^2 - \sum_{J\geq 0} O_J \star \phi^J \right\},
\end{equation}
where $h_{ij}(t,\xx)$ is the metric on each spatial slice and $\mathcal{N},\mathcal{N}^i$, $i=1,2,3$, are the lapse and shift
functions. We are using the notation
\begin{equation} \label{OJ}
 O_J = \sum_{n=0}^{n_J} (-1)^n \frac{\lambda_{J,n}}{\luv^{4-2n-J}} \Delta^n,
\end{equation}
with $\Delta$ the Laplacian defined with $h$,
\begin{equation}
 \Delta = h^{ij}\nabla^{(h)}_{i}\nabla^{(h)}_j,
\end{equation}
and $\luv$ a high energy scale which is left unspecified. It is implicit that $\Delta^0=1$.
The $\star$ encodes in a condensed notation the action of the various operators on 
$J$ copies of the field.
In particular, it is intended that for each $J$ one has to include all the operators obtained by all the possible independent combinations of $\Delta$ and $\phi$. For instance (taking the same example of \cite{FieldLifshitz}),
\begin{equation*}
 \Delta^3 \star \phi^3 = c_1 (\Delta \phi)^3 + c_2 \phi (\Delta \phi) (\Delta^2 \phi) + c_3 (\Delta^3\phi) \phi,
\end{equation*}
where $c_1,c_2,c_3$ are additional arbitrary constants.

In the expression for each $O_J$, given in \eqref{OJ}, the sum includes at most $n_J$ terms. The integers $n_J$ are fixed by renormalizability conditions.
In fact, it turns out that, in order to have power counting renormalizability, it is required that:
\begin{equation}
 n_J = \max \left\{ n \in \mathbb {N} \,| \, n\leq \frac{z+d}{2} + \frac{z-d}{4} J \right\},
\end{equation}
where $z$ is the dynamical critical exponent encoding the anisotropic scaling and $d$ is the number of spatial
dimensions (and hence in our case $d=3$). Notice that $n_2$ is just the integer part of $z$, whatever is the number
of spatial dimensions.

Without entering into many details, it is clear that, in perturbation theory around $\phi=0$, the properties of
propagation are encoded in the kinetic term alone, \ie in the part of the Lagrangian quadratic in the field, involving
therefore the terms corresponding to $J=2$. 

In doing this, we are discarding the terms corresponding to $J=0$ and $J=1$. 
The $J=0$ term is just a constant shift of the Lagrangian, irrelevant for the equations of motion of the matter fields (but contributing to the gravitational equations),
while $J=1$ contains only two possible kind of terms, namely $\beta\phi$ and $\gamma_n \Delta^n \phi$. While the latter are irrelevant boundary terms, the former would have the effect of shifting the minimum energy state from $\phi=0$ to $\phi=\phi_*$. For simplicity we assume that $\phi=0$ is a local minimum of the potential and hence $\beta=0$. 

Of course, if the potential is minimized by non-vanishing constant values of $\phi$, the propagation of perturbations will receive contributions also from terms with $J>2$. Nevertheless, this would not change the outcome of the discussion, while the restriction we are considering is just making the algebra more transparent.
Therefore, the non-quadratic terms, at this level, can be seen as mere (derivative) interactions, and we will neglect them.

By fixing the gauge to be $\mathcal{N}=1,\mathcal{N}^i=0$, we can reduce the analysis to the following Lagrangian
\begin{equation}
 L_0 = \frac{1}{2} \left( (\partial_t\phi)^2 -  h^{ij}\partial_i\phi \partial_j \phi -
\sum_{n=2}^{n_2} \frac{\alpha_n} {\luv^{2n-2}} \phi \Delta^n \phi -m^2 \phi^2\right),
\end{equation}
and the consequent wave equation. Here we have used $\alpha_{n}$ instead of $(-1)^n\lambda_{2,n}$ to simplify the notation. Furthermore, without loss of generality,
we have fixed $\lambda_{2,1}=1$, and we have defined:
\begin{equation}
 m^2 = \lambda_{2,0} \luv^2.
\end{equation}
Of course, for $z=1$ we recover the familiar relativistic dispersion relation. In the following, then, we will
assume $z\geq 2$ (and, correspondingly, $n_2\geq 2$).

The eikonal approximation is the key technical tool needed to grasp the geometrical meaning of the wave equation.
The analysis can be found, for instance, in \cite{Suyama:2009vy, Capasso:2009fh}. First, the field is represented as
\begin{equation}
 \phi(t,\xx) = A(t,\xx) e^{-iS(t,\xx)},
\end{equation}
where it is implicit that one has to take just the real or imaginary part, in the case of real scalar field $\phi$.

In the limit of very short wavelengths, \ie the geometric optics limit, the wave equation reduces to the so-called  eikonal equation, effectively describing the propagation of wave-fronts, encoded into the (rapidly varying) eikonal function $S$. In this case, it is given by
\begin{equation}
\left(\frac{\partial S}{\partial t}\right)^2 -
 \left( h^{ij}\frac{\partial S}{\partial x^i}
\frac{\partial S}{\partial x^j}\right) - \sum_{n=2}^{n_2} \frac{\alpha_n}{\luv ^{2n-2}}\left(-
 h^{ij}\frac{\partial S}{\partial x^i}
\frac{\partial S}{\partial x^j}\right)^n -m^2 = 0.
\end{equation}

The geometrical interpretation is easily given once it is recognized that the eikonal equation is
the Hamilton--Jacobi equation for a (fictitious) point particle (moving along the ray, in geometrical optics) whose
 action (in the Hamiltonian formalism) is:
\begin{equation}
 I = \int \left\{ \dot{x}^{\mu} p_\mu - \lambda(\tau) \left( G^2(p_0,p_i)-m^2 \right)\right\} d\tau,
\label{ActionHam} 
\end{equation}
with
\begin{equation}
G^2(p_0,p_i) = p_0^2 -  h^{ij}p_ip_j -
 \sum_{n=2}^{n_2} \frac{\alpha_n }{\luv ^{2n-2}} (h^{ij}p_ip_j)^n,
\end{equation}
and $\lambda$ a Lagrange multiplier used to enforce the mass-shell constraint.

From its form, it is manifest that the propagation of the ray involves the properties of the $3-$geometry
of each slice $\Sigma$. However, as we shall see, it is possible to introduce a notion of four dimensional geometry, \ie it is possible to show that the rays are just geodesics of a suitably defined four dimensional geometrical
structure. This has been done in the ``flat'' case in \cite{GLS}. There the outcome was that the 
particles were moving along the geodesics of (flat pseudo-)Finsler metrics. 

Finsler spaces are a class of spaces where the metric properties are encoded into a norm,
\ie the length of an arc of a curve $\gamma_{AB}$ is given by:
\begin{equation}
 \ell(\gamma_{AB}) = \int_{a}^b d\tau F(x,\dot{x}),
\end{equation}
where $F(x,\dot{x})$ is such that, at each point $m$ with coordinates $x$, the function $F(x,-):T_m 
\mathcal{M} \rightarrow \mathbb{R}$, defined over the tangent space $T_m\mathcal{M}$ at a given point $m$, 
 does obey the axioms defining a norm
(see, for instance, \cite{baochernshen} for a comprehensive exposition, while for a very basic discussion, see \cite{GLS}, section II). Clearly, Riemannian geometry is a particular case of Finsler geometry, where the norm is induced by the metric tensor
\begin{equation}
 F_{\mathrm{Riem}}(x,\dot{x}) = \sqrt{g_{ij}(x)\dot{x}^i\dot{x}^j}.
\end{equation}

To show that the rays in Ho\v{r}ava--Lifshitz scenarios are just geodesics of some Finsler 
spacetime, technically, one should proceed with a Legendre transform of the action \eqref{ActionHam}. Of
course, this will be possible if and only if the relationship between ``velocities'' and momenta, 
\begin{equation}
 \dot{x}^{\mu}= \lambda \frac{\partial G^2 }{\partial p_\mu},
\end{equation}
is an invertible map. When it is possible to go to the Lagrangian formalism, after the elimination of 
the Lagrange multiplier $\lambda$, it is immediate to realize that the action has the form:
\begin{equation}
 I=m \int d\tau F_{m}(x,\dot{x},\{\alpha\},\luv),
\end{equation}
where $F_{m}$ is a norm.

This conclusion can be inferred immediately from the symmetries 
of the action \eqref{ActionHam}. 
As one can show, this action is reparametrization invariant 
(provided that $\lambda \rightarrow (d\tau/d\tau') \lambda $).
Hence, since this property is never touched in the process of Legendre-transforming, it must be present 
also in the Lagrangian formalism. However, this is possible if and only if the Lagrangian is a 
homogeneous function of the velocity,
\begin{equation}
 F(x,s \dot{x}) = s F(x,\dot{x}), \qquad s\geq 0.
\end{equation}
This of course is not enough to prove that $F$ is a norm: one should check that all the axioms are 
satisfied. However, given that this structure would describe spacetime, and not just space, one should 
provide a definition of norms for Lorentzian signature. This is still an open problem. A rather 
conservative approach is to define a pseudo-Finsler structure a pair $(\mathcal{M},F)$ if $F$ is a
homogeneous function of the vector argument and if the tensor:
\begin{equation}
 g^{\mathrm{Finsler}}_{\mu\nu}(x,v) = \frac{1}{2} \frac{\partial F^2}{\partial v^{\mu}\partial v^{\nu}},
\end{equation}
called the Finsler metric tensor\footnote{Notice that in the case of Riemannian geometry, 
$F^2(x,v)=a_{\mu\nu}v^\mu v^\nu$, whence $g^{\mathrm{Finsler}}_{\mu\nu}(v)=a_{\mu\nu}$.} is non-degenerate with signature $(-+++)$ (or $(+---)$, according to the conventions) \cite{Beem}.
Again, for the technical details and comments see \cite{GLS}.

This routine can be applied without major changes also in the case of Hor\v{a}va--Lifshitz scenarios. 
The 
difference will be that now the geometrical structure defined by the given dispersion relation will be 
curved, in general. The main outcome, however, is unchanged: particles obeying mass-shell relations like the one in
\eqref{ActionHam} are propagating on geodesics of suitably defined Finsler structures. Alternatively,
in the geometric-optic limit of the wave equations in Ho\v{r}ava--Lifshitz scenarios, the rays are 
geodesics of suitably defined (four dimensional) pseudo-Finsler structures. 

Some specifications are in order. As in the case of MDR discussed in \cite{GLS}, the norm depends on the 
mass of the particle. This is somehow unavoidable, even in the case in which the coefficients $\alpha_n$ 
are particle-independent. See also \cite{LoHiggs} for a discussion of some related aspects for 
spontaneously broken gauge symmetries. This means that we cannot globally replace the $3+1$ formalism with
a single Finslerian framework, at least at this stage. 

Furthermore, there is another source of difficulties for such a perspective.
In the case of higher spins a more involved normal modes analysis can be done, along the same lines
briefly sketched for the case of the scalar field. For a detailed
 discussion see \cite{NormalModes}. 
In the most general case, one should expect multi-refringence,
\ie each polarization of the field does propagate on a different geometrical structure. The emerging geometrical
structures are still non-Riemannian, and Finsler geometry is still playing an important role.
However, for a critical discussion about the possibility of describing multi-refringence by means of a single
Finsler structure, see, for instance, \cite{Skakala}.

Despite these issues, which require further investigations, the main lesson is clear:
in Ho\v{r}ava--Lifshitz scenarios, the description of the propagation of waves can be given 
in terms of curved four-dimensional pseudo-Finsler structures.
This is just the generalization of what has been discussed in \cite{GLS} in the case of flat spacetime 
geometries. Concretely, one should expect that the higher is the energy of the particle, the larger will
be the deviation from the geodesic motion determined by the low-energy four dimensional metric $g_{\mu\nu}$ \cite{Capasso:2009fh}. 
Nonetheless, the motion of particles, dual to the propagation of waves, is still geodesic, once it
is realized that the metric structure is (pseudo-)Finslerian, rather than (pseudo-)Riemannian.

\vspace{1cm}

{\bf Acknowledgments} I want to thank F. Girelli and S. Liberati for useful comments.

\end{document}